# Numerical investigations of the minimum-B effect in Electron Cyclotron Resonance Ion Source

V. Mironov, S. Bogomolov, A. Bondarchenko, A. Efremov, V. Loginov, D. Pugachev

*Joint Institute for Nuclear Research, Flerov Laboratory of Nuclear Reactions,*
*Dubna, Moscow Reg. 141980, Russia*
E-mail: vemironov@jinr.ru

Abstract: The three-dimensional particle-in-cell model NAM-ECRIS is used for investigation of how the DECRIS-PM Electron Cyclotron Resonance Ion Source is reacting to changes in the source magnetic configuration. The accent is made on changes in the magnetic field at the magnetic trap center, the minimum-B value. It is calculated that the optimal normalized value of the field is ~0.8, close to the experimental observations. The reasons for existence of the optimum are discussed. It is observed that the electron energies are increasing with the increased minimum-B values due to enhanced confinement of the energetic electrons in the plasma. Bumps in energy spectra of the radially lost electrons are observed and explained to be due to nonadiabatic losses of electrons.

## Introduction.

The Electron Cyclotron Resonance Ion Sources (ECRIS) are versatile tools aimed to produce intense beams of highly charged ions for variety of applications. Plasma in the source is confined in the minimum-B open magnetic trap and is heated by resonant absorption of intense microwave radiation by electrons. Ions are produced by sequential ionization by electron impact and reach the high charge states if electron and ion confinement times are large enough. Typical electron densities in ECRIS are around $10^{12}$-$10^{13}$ cm$^{-3}$, the ion life time is around 1 ms, which is sufficient for production of the ion beams with the currents in the mA range for such ions as Ar$^{8+}$ and U$^{35+}$.

The sources operate stably and produce the maximized ion currents if the source parameters are tuned properly. Many factors affect the source performance, such as the gas flow and composition, power and frequency of the injected microwaves, wall conditions etc. Magnetic field configuration is of the special importance because it directly influences the main parameters of the source plasma. The sources are designed by following the empirically established scaling laws [1], which request that the hexapole magnetic field at the source chamber walls ($B_{rad}$) should be at least twice larger of the resonant magnetic field ($B_{ECR}$), $B_{rad,n} = B_{rad}/B_{ECR} \approx 2$, the field at the extraction $B_{ext} \approx B_{rad}$, the field at injection $B_{inj,n} = B_{inj}/B_{ECR} \approx (3\text{-}4)$, and the minimal field at the source axis $B_{min,n} = B_{min}/B_{ECR} \approx (0.7\text{-}0.8)$.

The requirement for the $B_{min,n}$ value follows from the experimental observations that the extracted currents for the medium charged ions are almost doubled when increasing the minimum-B value from ~0.5 to ~0.8 with keeping other parameters the same [2]. For the $B_{min,n}$ value above the optimum, the currents either saturate or decrease. Also, it is measured that the bremsstrahlung intensity and spectral temperature are steadily increasing with the $B_{min,n}$ value [3]. The reasons for the boosted heating of electrons in the plasma are explained by variations in the magnetic field longitudinal gradient at the ECR surface, which affects the electron heating rate [4].

It is observed that the extracted ion currents are unstable for the minimum-B values above the ~0.8, in correlation with bursts of the electron losses and the microwave emission from the plasma. These oscillations are attributed to be due to the electron cyclotron instability [5], which threshold depends on the electron energy anisotropy and electron density. Both these factors, the anisotropy and density, are supposed to increase with the minimum-B value; onset of the instability is conjectured to limit the source performance above the optimal minimum-B value.

Energies of the axially-lost electrons were measured in [6] and [7]. It was found that bump is formed in the energy spectra of the lost electron at energies of 200-300 keV for the 14 GHz AECRIS source [6] and 400-700 keV for the 18-GHz SUSI source [7]. Energy of the bump depends almost linearly on the minimum-B value for variations of $B_{min,n}$ in the range of (0.7 - 0.8). First attempts to explain the bump origin were connected to the relativistic shift in the electron cyclotron resonance with the electron energy [8]. For those electrons that resonate with microwaves close to the extraction aperture of the source, RF-induced diffusion in the velocity space can results in their enhanced loss rate thus forming the bump in the lost-electron energy spectra. However, such the explanation requests that the bump energy should be dependent on the extraction magnetic field, which is not confirmed by the measurements. Alternative approach involves a presence of low-frequency resonantly excited microwaves in the source chamber, which resonate with electrons close to the minimum-B position and push them into the loss-cone [9]. However, it is not confirmed experimentally that such waves are excited with intensities sufficient to noticeably affect the electron dynamics.

A lack of clear understanding of the minimum-B effects motivated us to numerically investigate the ECRIS parameters for a set of magnetic configurations with different $B_{min}$ values. Simulations were done by the NAM-ECRIS model [10] with the parameters of the DECRIS-PM source [11]. Results of the calculations confirm the main experimental observations. The paper is organized in the following way: first, the main features of the source are described; then, the results of the NAM-ECRIS calculations are given. In the last section, the results are extensively discussed to reveal the mechanisms behind the source responses to variations of the minimum-B values. It is found that the nonadiabatic losses of energetic electrons [12] play an important role for determining the source performance and mean electron energy in the ECRIS plasma.

### The DECRIS-PM source.

The source operates with injection of single-frequency 14.5 GHz microwaves with injected power of up to 500 W. The source chamber dimensions are 23 cm in length and 7 cm in internal diameter. The ions are extracted through extraction aperture of 1 cm in diameter at the extraction side of the source. At the injection flange, there are installed the waveguide for injection of the microwaves, the aluminum biased electrode with the diameter of 3.0 cm, the gas injection port and the port for the micro-oven installation. Both ports are located off-axis at the radius of 2.3 cm. The flange is perforated for better pumping of the source chamber, the holes cover 25% of the surface not shielded by the biased electrode. The chamber is made of polished stainless steel. The biased electrode voltage is up to -1250 V, the extraction voltage is up to 30 kV; the optimized bias voltage is around -250 V for highly charged argon ion production.

The magnetic fields of the source are formed by a set of permanent magnets. The solenoidal component is produced by the ring-shaped magnets with the radial and axial magnetizations; the hexapole component is produced by the 24-segment Halbach magnet. The magnetic field on the axis is 1.34 T at the injection flange and 1.1 T at the extraction electrode. The hexapole field at the source chamber radial walls is 1.1 T. The microwave frequency of 14.5 GHz defines the resonant magnetic field of 0.518 T for cold electrons.

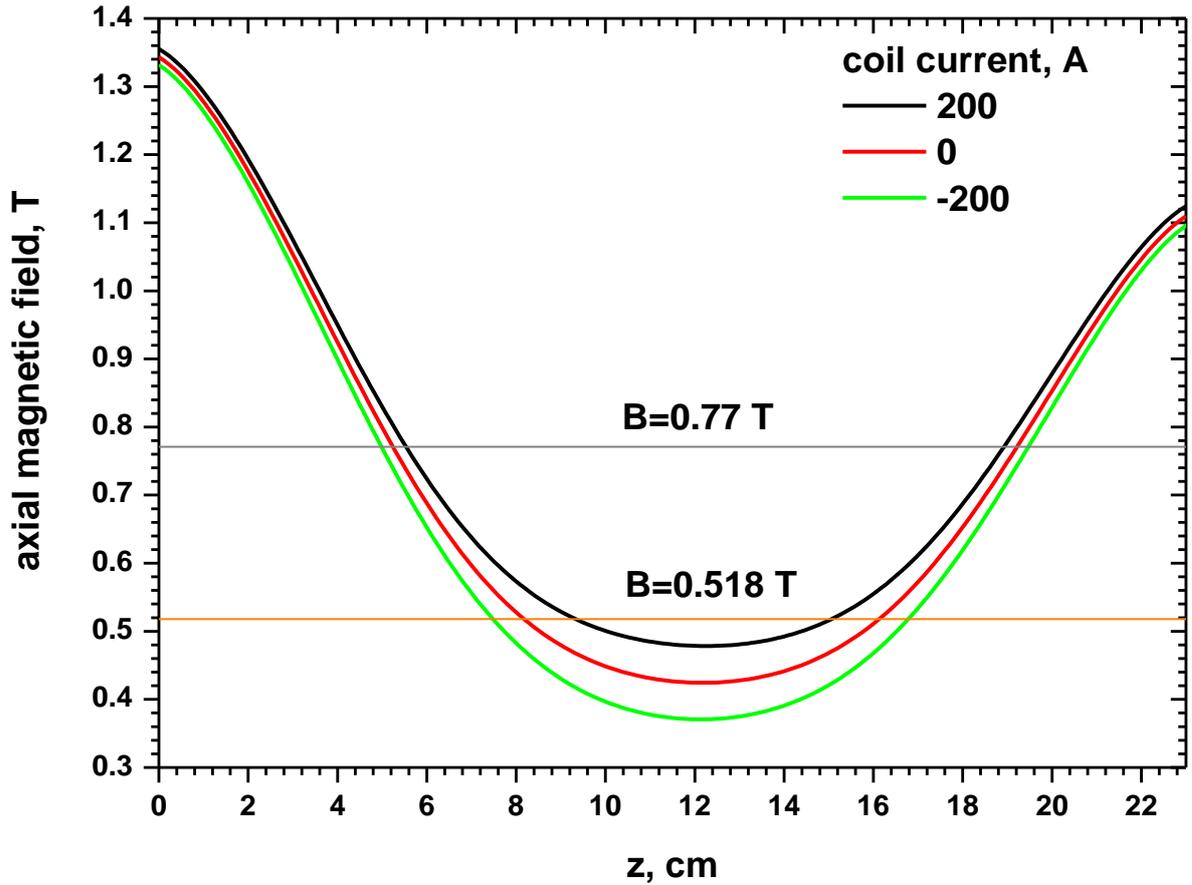

Fig.1. Axial magnetic field of DECRIS-PM for the currents in the correction coil of 200, 0 and -200 A. The cold ECR value (0.518 T) is shown as orange line, the relativistic ECR value (0.77 T) for 250 keV electrons is shown as grey line.

The source is equipped with the correction coil located close to the chamber center in between the hexapole and central magnetic ring; the coil inner diameter is 16.4 cm and the length is 10 cm. The maximal currents in the coil are ±300 A and the coil power consumption is up to 1.5 kW. Energization of the coil allows tuning the minimum-B value in the range of ±0.075 T with no substantial changes in the magnetic fields at the injection and extraction sides.

The magnetic fields of the source are calculated by the 2D magneto-static POISSON-SUPERFISH code [13] for three different currents in the correction coil. The field along the source axis is shown in Fig.1 for the currents of +200, 0 and -200 A. The positive currents increase the magnetic field along the axis close to the center, and the negative ones decrease it. The minimum-B values for the calculated profiles are 0.478, 0.424 and 0.370 T respectively, which corresponds to the normalized values of 0.92, 0.82 and 0.715. The resonant magnetic field of 0.518 T for cold electrons is shown in Fig.1 as the orange line. It is seen that the ECR zone length is decreasing for the larger minimum-B values, as well as the field longitudinal gradient. The relativistic value of ECR for 250 keV electrons (0.77 T) is shown as grey line; changes in the ECR zone length and in the magnetic field gradient are not significant for different minimum-B values.

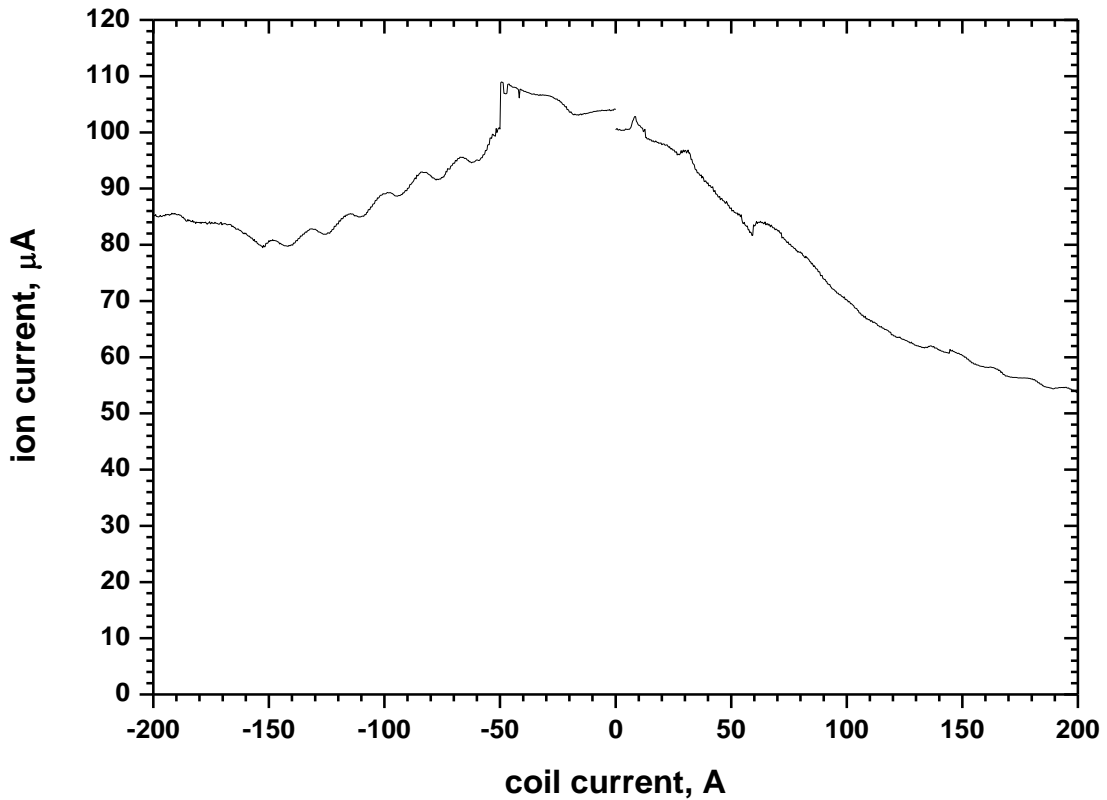

Fig.2. Extracted currents of $Ar^{8+}$ ions as the function of the correction coil current.

The experimentally measured extracted currents of $Ar^{8+}$ ions are shown in Fig.2 as the function of the correction coil current. The source was optimized during the measurements for the current stability and the ion currents were relatively small. The ion current is maximized at the coil current of ~ (-50) A and decreases for the positive coil currents; the optimal minimum-B value for this source is close to 0.8. After optimizing the gas flow and the injected microwave power, the $Ar^{8+}$ ion beam of (0.5-0.6) mA can be extracted, which is the typical value for the sources of this generation. Gas-mixing of argon and helium allows further increasing the currents of $Ar^{8+}$ argon ions up to the level of (0.8-1) mA.

The source parameters are used for numerical simulations of the ECRIS plasma by the NAM-ECRIS model. The calculations are done for the set of the magnetic field profiles calculated for the correction coil currents shown in Fig.1.

### The NAM-ECRIS model.

The numerical model is extensively described elsewhere [10]. The model is a set of two iteratively running 3D particle-in-cell modules that separately simulate the electron and ion dynamics in the plasma. Additionally, the COMSOL Multiphysics® software [14] is used for simulations of the microwave propagation in the magnetized ECRIS plasma. The COMSOL's model provides the spatial and phase distributions of the microwave electric field amplitude in the source chamber by using the 3D electron density and the magnetic field distributions obtained by the electron module. In turn, the electric field amplitudes and phases are used by the electron module to simulate the cyclotron resonance heating of electrons. Electron scattering in collisions with the plasma electron and ion components is treated by using the corresponding spatial distributions from the previous runs of the electron and ion modules. The electron module calculates the energy and spatial distributions of electrons in the plasma, as well as the globally defined electron life time. The ion module uses this information to trace the ion movement and processes of ionization, charge-exchange and neutralization on the walls.

Calculations begin with assuming an arbitrarily selected seed electron spatial distribution (usually, with setting the uniform electron density inside the ECR zone) and are continued in the iterative way until the converged solution is obtained. During the iterations, plasma is characterized by the total electric charge of electrons in the plasma, which is fixed at the level of 6.7 µC for all investigated minimum-B configurations. The converged solution provides the main plasma parameters such as the plasma density and the electron energy distribution functions, the fluxes of the injected material and the extracted ion currents. Also, plasma potential distributions are obtained in the model, as well as the electron and ion life times.

The described calculations are done for argon with no mixing with other gases. The microwave injected power is fixed at 500 W. The biased electrode voltage is set to -250 V, the extraction (i.e., retarding the electron fluxes through the extraction aperture) voltage is 20 kV. We assume that the argon is leaving the source chamber if atoms and ions are hitting the extraction aperture of 1-cm in diameter. Also, if particles hit the injection flange in the regions not shielded by the biased electrode with 3-cm in diameter, then they are considered to be lost with a probability of 25% in accordance with the total relative area of the pumping holes. Argon atoms are injected with the angular distribution corresponding to the cosine-law through the port at the injection flange. The port separation from the source axis is 2-cm; it is located in between the plasma star arms.

To reach a converged solution, 5-6 iterations are needed for each magnetic field configuration. Each iteration run lasts around 24 hours on a quad-core 3.2-GHz PC. The electrons are traced for ~1 ms of the physical time with the time step of 10 ps; normally, we use $10^3$ computational particles on the relatively coarse 65×65×64 cartesian mesh in x-, y- and z-directions respectively. The ion module uses $4×10^5$ computational particles on the same mesh as the electron module, the particles are traced for ~10 ms of the physical time with the time step of 0.1 µs.

### Results.

The converged spatial distributions of electron density are shown in Fig.3 along the source axis (a) and in the transversal x-direction at z=12.1 cm, which corresponds to the minimum-B position at the axis (b). The distributions are shown for the correction coil currents of 200, 0 and -200 A, that is, for the normalized $B_{min,n}$ values equal to 0.92, 0.82 and 0.715 (see Fig.1).

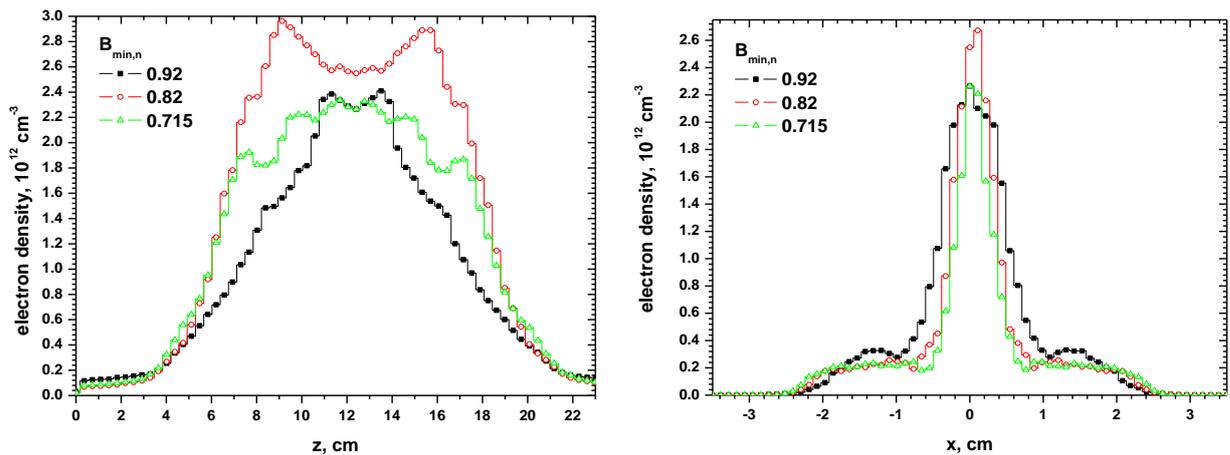

Fig.3. The electron density along the source axis (a) and in the transversal direction (b) for the coil currents of 200, 0 and -200 A corresponding to the normalized minimum-B values of 0.92, 0.82 and 0.715.

The electron densities at the source axis are in the range of (2.0-2.6) ×$10^{12}$ cm$^{-3}$, being maximized for the intermediate minimum-B value of 0.82. Longitudinally, plasma is more compact for the large $B_{min,n}$ value. The dense plasma core is formed along the source axis surrounded with the relatively dilute and hot plasma halo. The transversal size of the plasma (standard deviation of Gaussian fit to the density profile, σ) is increasing with the $B_{min,n}$, σ=2.1, 2.4 and 3.9 mm for the $B_{min,n}$=0.715, 0.82 and 0.92 respectively.

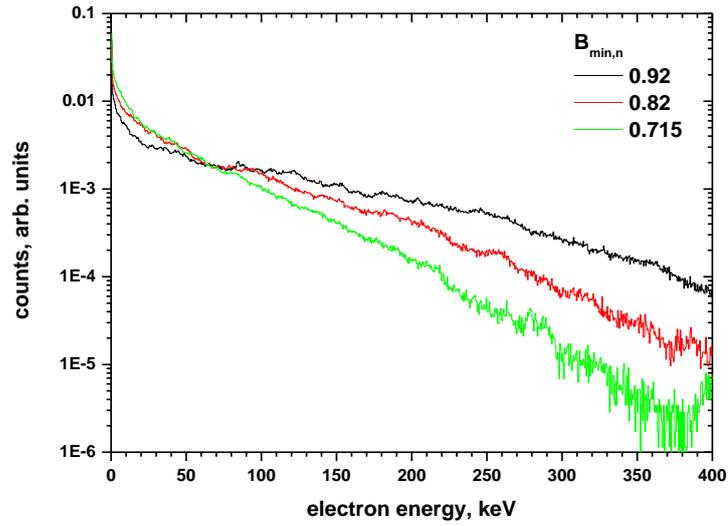

Fig.4. Energy distributions of the trapped electrons for different minimum-B values.

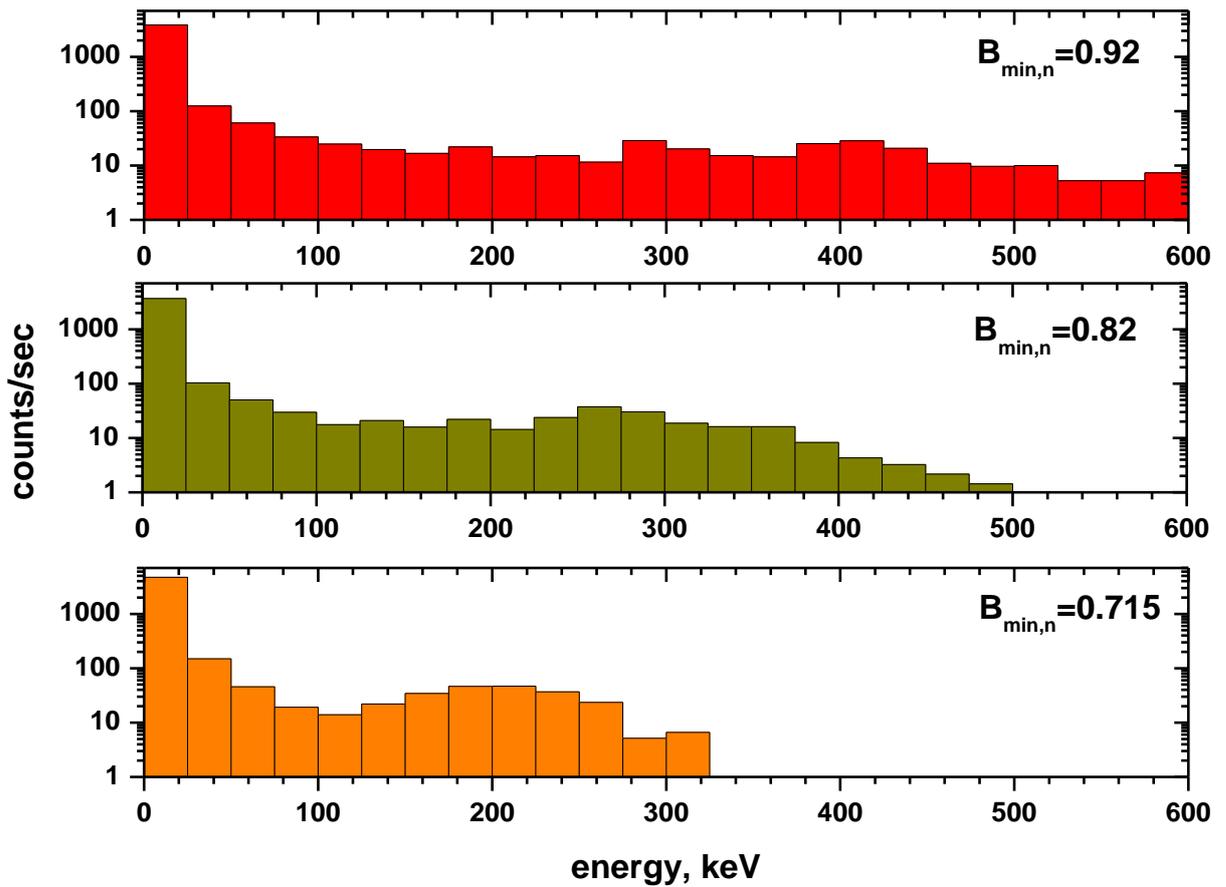

Fig.5. Energy distributions of the lost electrons for different minimum-B values.

The electron energy distributions of the trapped electrons integrated over the plasma volume are shown in Fig.4 for the investigated set of the $B_{min,n}$ values. The distributions are normalized to same total numbers of counts.

The distributions are fitted with the exponentially decaying curve for the energies above 20 keV. The fit slopes are 50, 81 and 126 keV for the $B_{min,n}$=0.715, 0.82 and 0.92 respectively. Following the increase in the electron spectral temperatures, the electron mean energies are 39, 59 and 95 keV. The larger is the minimum-B value, the more energetic are electrons inside the plasma. The calculations are consistent with the experimental observations [15], which confirm the electron spectral temperature increase with minimum-B.

Energy distributions of the lost electrons differ from the energies of the trapped particles. Distributions of the lost electron energies are shown in Fig.5 for the investigated set of minimum-B values. The mean energies of the lost electrons are 14, 23 and 35 keV for the minimum-B values of 0.715, 0.82 and 0.92 respectively, around 35% of the trapped electron mean energies. This is consistent with the experimentally observed differences between the X-ray spectra of bremsstrahlung measured axially and radially [16]. The axial spectra include emission from electrons that are lost at the biased electrode and at the injection flange, and therefore mostly reflect the lost electron energies. The radial emission is mostly due to electron-ion scattering and is defined by the electrons that are trapped in the plasma.

Bumps are seen at the distribution tails; the bump energy is shifting to the larger values when increasing the $B_{min,n}$ values. For the lowest minimum-B, the bump's position is ~200 keV and it reaches 400 keV for the $B_{min,n}$=0.92. Electrons are preferentially lost at the radial walls of the source chamber, electron losses toward the injection and extraction flanges are less than 15% of the total losses. For the statistical reasons, we are not able to resolve the spatial dependencies of the lost electron energy distributions. Authors of [6] measured the lost electron energies for those particles that escape the source axially. Also, they did not apply the extraction voltage to the source body during their measurements, which strongly influences the plasma spatial distributions [8]. Therefore, spectra in Fig.5 cannot be directly connected to the measured distributions. However, the spectra variations with minimum-B values closely resemble the experimentally observed tendencies of [6].

Spatial distributions of the microwave electric field amplitude are not changed substantially with the variations of the minimum-B value. The wave amplitudes (not resolved in their polarizations) are at the level of 100 V/cm close to the corresponding ECR positions, decreasing inside the cold ECR zones. The gas flow into the source chamber, which is needed to reach the converged plasma parameters with the fixed total content of electrons, only slightly varies with the minimum-B value, increasing from 1.8 to 2.0 particle-mA for $B_{min,n}$=0.715 and 0.92 respectively. The electron life time is maximal (0.25 ms) for $B_{min,n}$=0.82, being substantially lower (0.2 ms) for $B_{min,n}$=0.715 and almost the same (0.23 ms) for $B_{min,n}$=0.92. The calculated plasma potential is around 0.2 V at the plasma center, with the weak dependence on the minimum-B value; the potential dip is ~0.02 V. The ion temperature inside the plasma is ~0.15 eV for the argon charge states 8+ and higher.

The ion extraction efficiency is defined as the ratio between the flux through the extraction aperture of ion in the given charge state and the total flux of these ions toward the walls. The extraction efficiency of $Ar^{8+}$ ions is ~0.3 for the relatively low minimum-B values and it decreases to 0.2 for the $B_{min,n}$=0.92 due to stronger transversal transport of ions during their movement from the dense parts of the plasma toward the extraction.

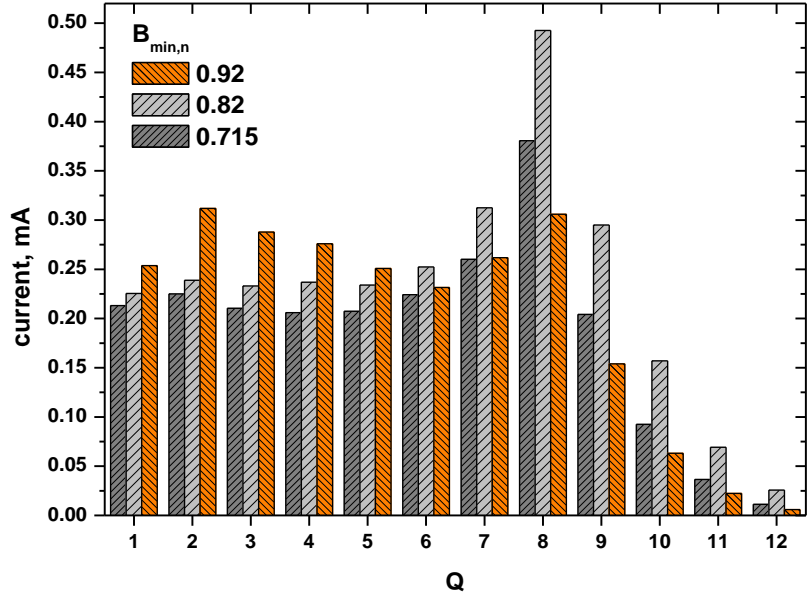

Fig.6. Charge-state distributions of the extracted argon ions for the different minimum-B values.

The calculated charge state distributions of the extracted ions are shown in Fig.6. Currents of the ions with the charge state above 7+ are maximized for the $B_{min,n}$=0.82, that is, for the non-energized correction coil. The tendency is the same as for the experimental data presented in Fig.2. The main factors that define the extracted ion current variations are changes in the electron life times and in the ionization rates: for energies much higher of the ionization potentials of ions, the hotter are electrons, the lower are the rates. For the relatively cold electrons at $B_{min,n}$=0.715, the rates are highest, but this is counteracted with the relatively low electron (ion) life time (mostly due to losses of energetic electrons seen in Fig.5) and, therefore, production of the highly charged ions is less efficient in these conditions. With the large minimum-B value, the ionization rates are low because electrons are over-heated, and, for the comparable electron densities and life time, this causes the calculated decrease in the extracted ion currents. We see that the key factor in the minimum-B effect is the noticeably better electron energy confinement and large electron energies for large $B_{min,n}$ values. Reasons for increase in electron mean energies are discussed in the next section.

## Discussion.

The ECRIS plasma can be conventionally divided in two different parts: the dense core plasma and the hot diluted halo. The core plasma is formed by reflections of the relatively cold electrons from the potential barriers formed close to the biased electrode and in the extraction aperture [8]. The relative importance of the retarding potentials is decreasing with electron energization by microwaves, and the hot electrons are trapped by the magnetic mirrors far from the walls. The halo electrons are moving such that some of their turning points are located near the radial walls, with smaller mirror ratio compared to the core.

Trajectories of the electrons in the source chamber are visualized in Fig.7 in the transversal and longitudinal planes for two magnetic field configurations with the minimum-B values of 0.715 and 0.92. The transversal (x-y) slice is located at z-position of 12.1±0.1 cm close to the minimum-B position, the longitudinal slice in the y-z plane corresponds to span δx of ±1 mm. The brighter is a pixel color in the pictures, the more electrons crossed this point over the acquisition time. The source chamber walls are shown in the pictures as the black lines.

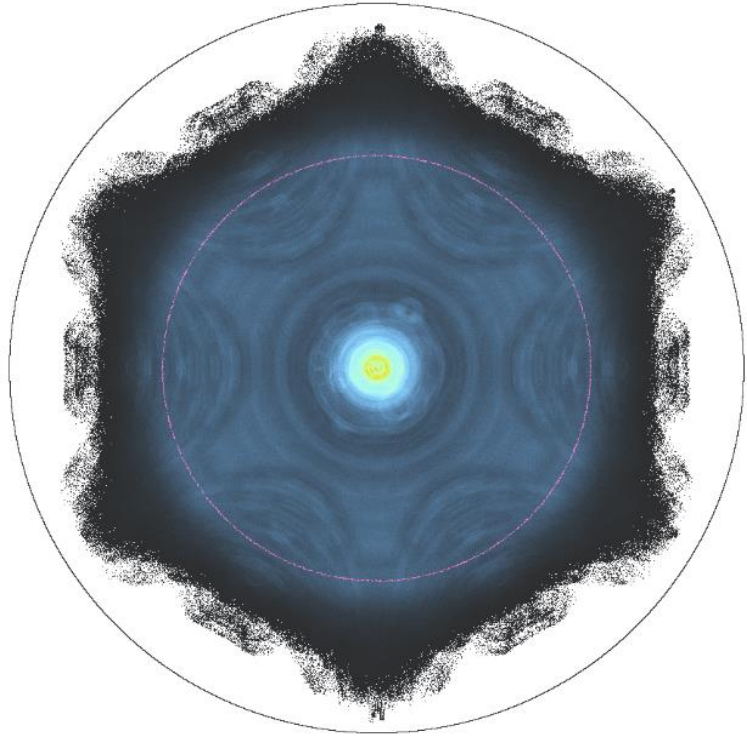

(a) $B_{min,n}=0.715$

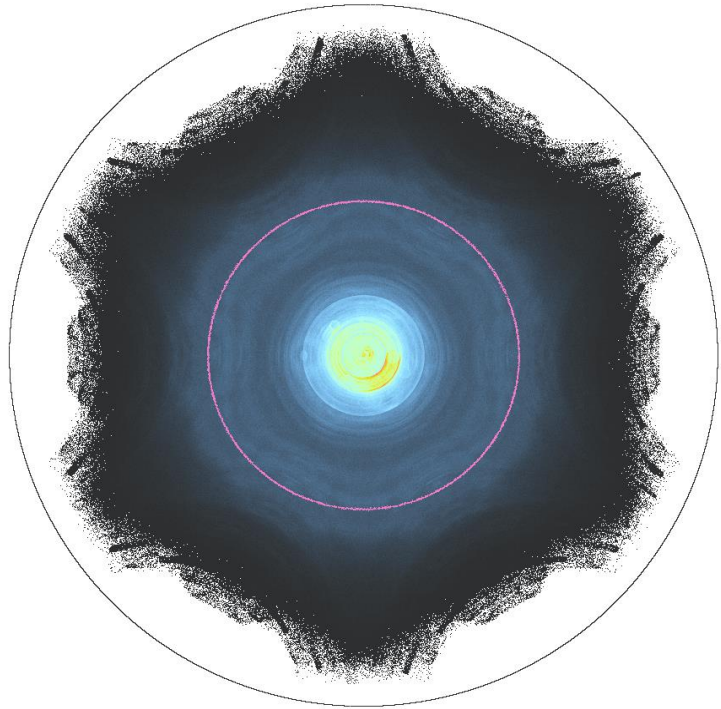

(b) $B_{min,n}=0.92$

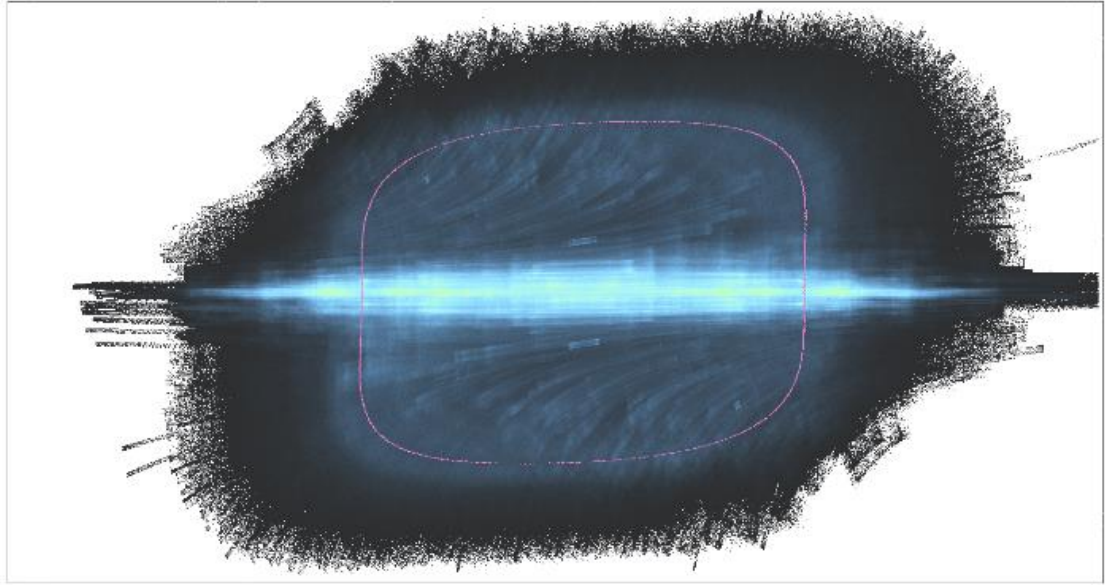

(c) $B_{min,n}=0.715$

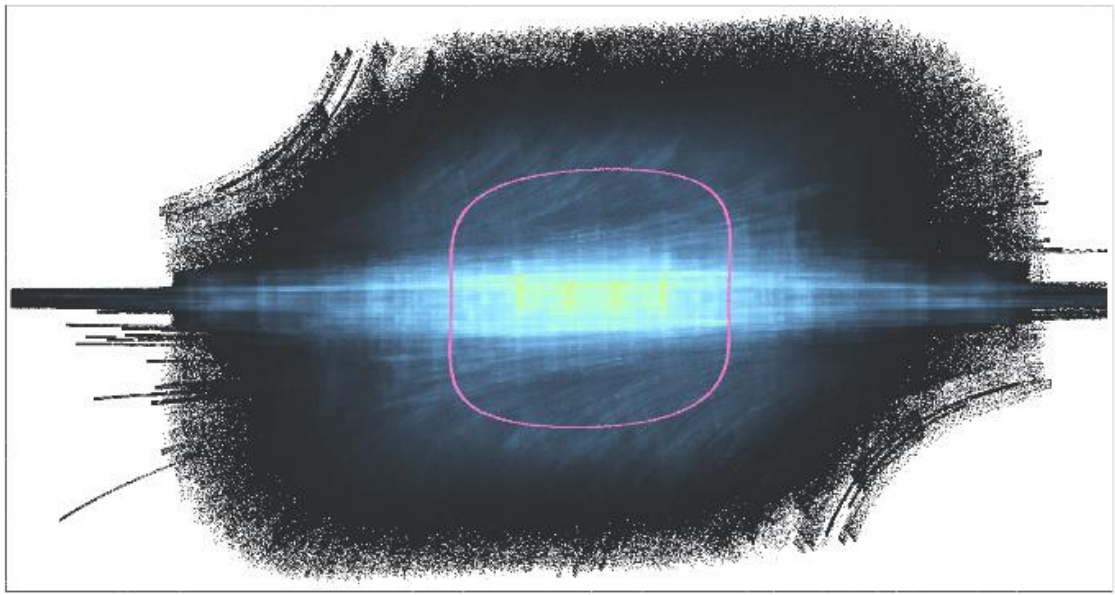

(d) $B_{min,n}=0.92$

Fig.7. Trajectories of electrons in the transversal (a) and (b) and longitudinal planes (c)-(d) for the normalized minimum-B values of 0.715 and 0.92.

The ECR surface for cold electrons is shown with pink line in the plots. The zone is more compact in longitudinal and transverse directions for the magnetic field configuration with the large minimum-B, and, on average, electrons are moving in the larger magnetic fields. The electrostatically confined electrons are seen close to the source axis in the longitudinal planes. The core electrons are forming the central bright spots. Transversal size of the core is noticeably smaller for the small minimum-B value.

Magnetically trapped electrons in the halo occupy the volume with sharp boundaries defined by the value of smallest magnetic field at the radial walls. The closest approach of this volume to the walls is at six leak points, where the magnetic field at the radial walls reaches the minimum. The electron fluxes to the walls are preferentially localized at these points, which explains the experimentally observed wall damage patterns [17]. The magnetic fields at three leak points close to the extraction side of the source are 1.12 T and 1.07 T (the difference is 5%) for the $B_{min,n}$ of 0.92 and 0.715 respectively.

The magnetic field in ECRIS is a combination of the solenoidal and the multipole (typically, hexapole) components. For the hexapole component, the magnetic field is zero on the source axis and increases with the radius (r) quadratically. Transversally, the solenoidal component decreases at the source central regions and increases close to the field maxima. The result of summation of two components is that close to the source axis the magnetic field is essentially solenoidal with the concave magnetic field lines; the lines start to be convex at larger distances from the axis. The leak points at the radial walls are the places where the radial component of the solenoidal component reaches its maximum in the direction opposite to the local radial hexapole field.

Electron movement in the magnetic field of the source is the Larmor rotation around magnetic field line, longitudinal bouncing for the mirror-trapped particles and relatively slow curvature drift perpendicular to the magnetic field lines. The electrons are diffusing both perpendicular and along the magnetic field lines due to collisions with other electrons and ions, resonant interaction with microwaves and nonadiabatic effects [18,19]. The latter channel is due to the fact that the particle magnetic moment $\mu = mv_\perp^2/2B$ is only an adiabatic invariant and it can change during the particle motion such that energy is redistributed between the transversal and longitudinal degrees of freedom with consequent escape of particle from the trap. A particle diffuses in the μ-space with the diffusion rate dependent on ratio between the Larmor radius and the characteristic length of the magnetic field variation (axial distance between the turning points, radius of curvature of the magnetic field line). The rate increases with the electron energy and nonadiabatic losses are especially pronounced for those particles that oscillate between the radial walls of the source, as it will be discussed later.

The curvature drift velocity is proportional to the $\frac{\mathbf{B} \times \nabla B}{B^3}$ term, where **B** is the magnetic field vector and $\nabla B$ is the magnetic field gradient. In Fig.8, we plot the azimuthal component of the curvature drift term in the plane of the source with x=0 for the minimum-B value of 0.715. The component is plotted in arbitrary units, with negative sign corresponding to the anti-clockwise azimuthal drift, i.e. in the direction of the electron Larmor rotation. The black lines separate regions with the negative and positive azimuthal drift velocities.

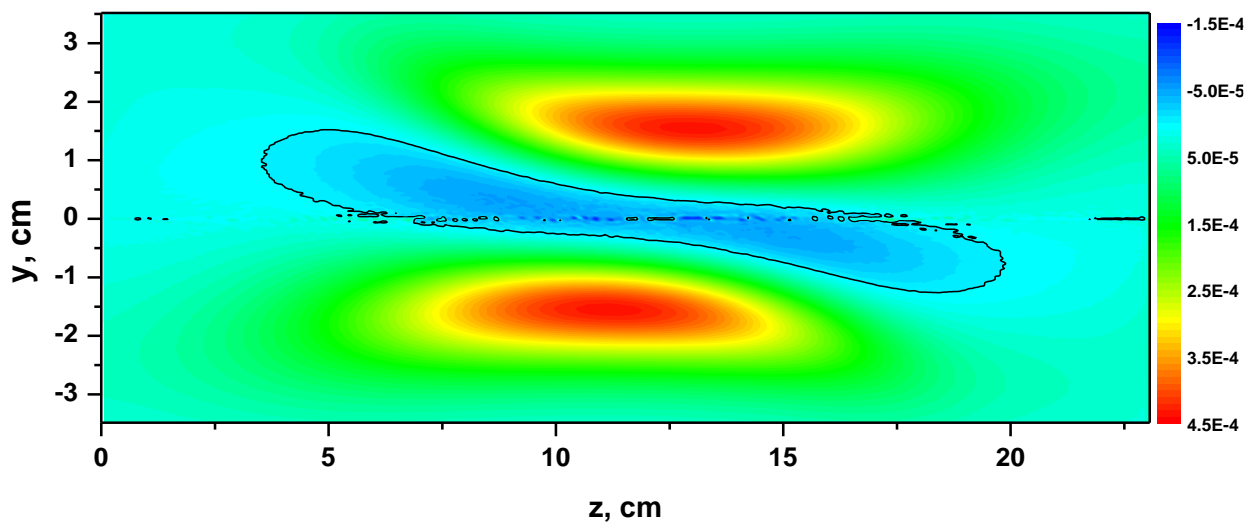

Fig.8. Azimuthal component of the curvature drift term (in arbitrary units) in the transversal plane (x=0) for the $B_{min,n}$=0.715.

For the larger minimum-B value of 0.92, the maximal local drift components are decreased by a factor of 1.5; transversal dimensions of the negative-drift region close to the source center are the same (Δy=±3.7 mm) for the small $B_{min,n}$=0.715 and the large $B_{min,n}$=0.92. For both minimum-B values, the local drift azimuthal components are maximized close to the source center at the radial displacements from the axis of around 1.5 cm.

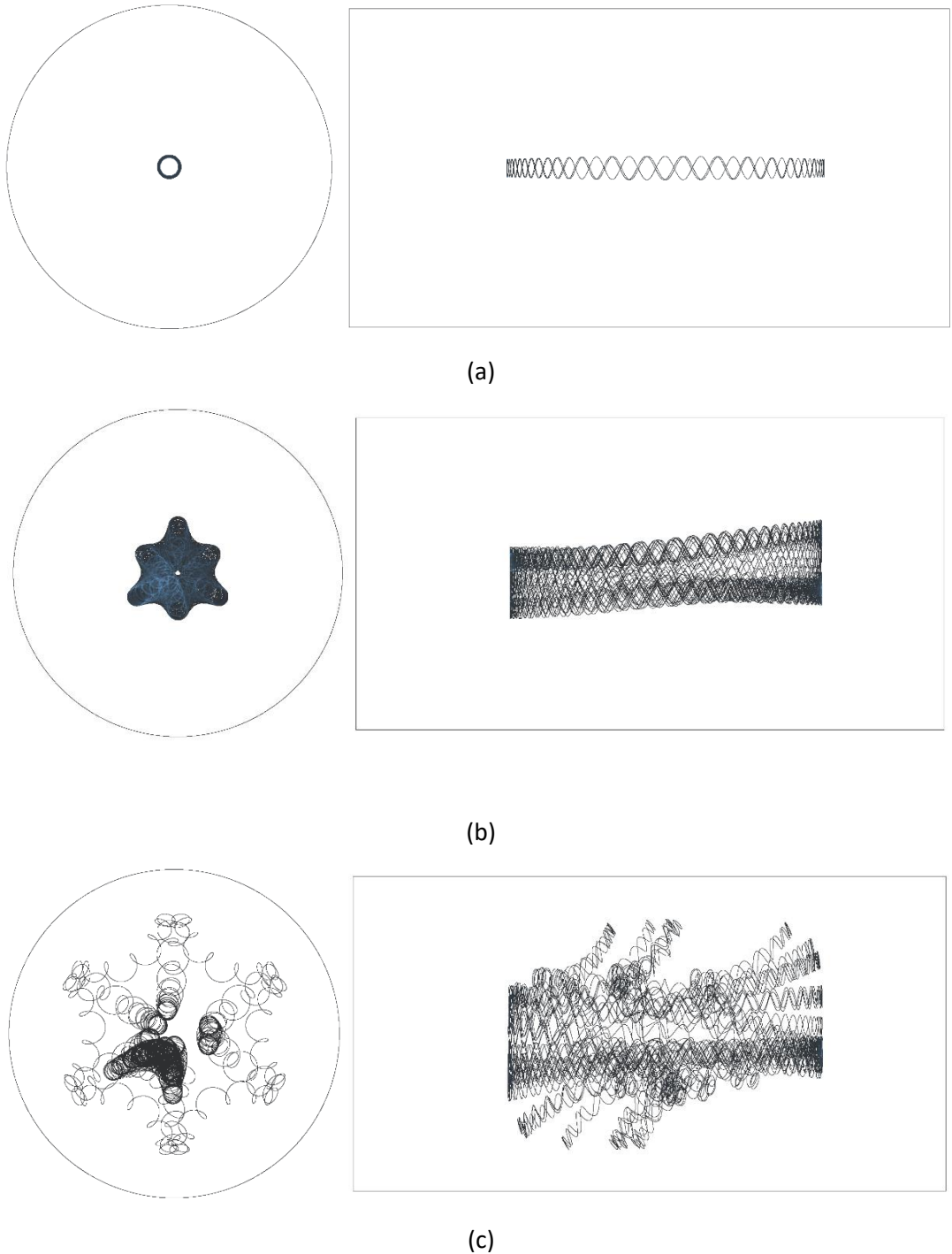

Fig.9. Trajectories of single electron in the transversal (x-y) and longitudinal (y-z) directions for the energy of electron of 125 keV and the minimum-B value of 0.715. The particle starting positions correspond to the initial guiding center displacement of 0, 2.35 and 2.45 mm for (a), (b) and (c) respectively.

During the electron movement in the source magnetic trap, direction of its global drift depends on integral of the field gradient orientation along the corresponding field lines. Close to the axis electron trajectories are concave; for the larger radial separations, the electron motion is influenced by the hexapole component of the magnetic field such that the trajectories are convex. Transition between two different types of motion is sharply dependent on the displacement of the electron orbit guiding center from the axis.

In Fig.9, we visualize the trajectories of a single electron with energy of 125 keV launched at three different positions along y-coordinate (in vertical direction), with zero displacement along x, and at the axial position z that corresponds to the local magnetic field of 0.645 T, the relativistic resonant value for electrons with such energy ($\gamma=1.245$). The particle is launched with zero longitudinal velocity – the initial position is located at the turning point. The collisional diffusion and interaction with microwaves are disabled when tracing the particle movement. All particle positions are shown, with no applied cuts along z- or x-coordinates. The trajectories are calculated for the minimum-B value of 0.715. Small time step for calculations is selected ($5\times10^{-15}$ sec) and it was checked by variations in the time step that the movement is numerically stable.

The first trajectory (a) begins with zero displacement of the guiding center from the axis – electron initial position along y-coordinate corresponds to the local Larmor radius ($r_L = \frac{m_e c \sqrt{\gamma^2-1}}{eB} = 1.96 \ mm$). The orbit for such electron in the x-y projection is a circle, with the electron moving back and forward along the source axis after reflections by magnetic mirror forces such that the guiding center is always on the axis. When initial position is shifted in the range of up to 2.35 mm (b), the electron motion remains to be preferentially axial, with the guiding center drift in direction of the Larmor rotation and with the limited displacement of orbits caused by the hexapole magnetic field component. Further displacement of the initial position from 2.35 to 2.45 mm (c) drastically changes the trajectory type: after few bounces along the source axis, the orbit starts to be radially directed, with the electron being reflected close to the radial walls and azimuthally drifting in the clock-wise direction. We see that there is a critical displacement of the electron orbit guiding center from the source axis, which switches the electron motion from axial to radial mode resulting in electron transport out of the core.

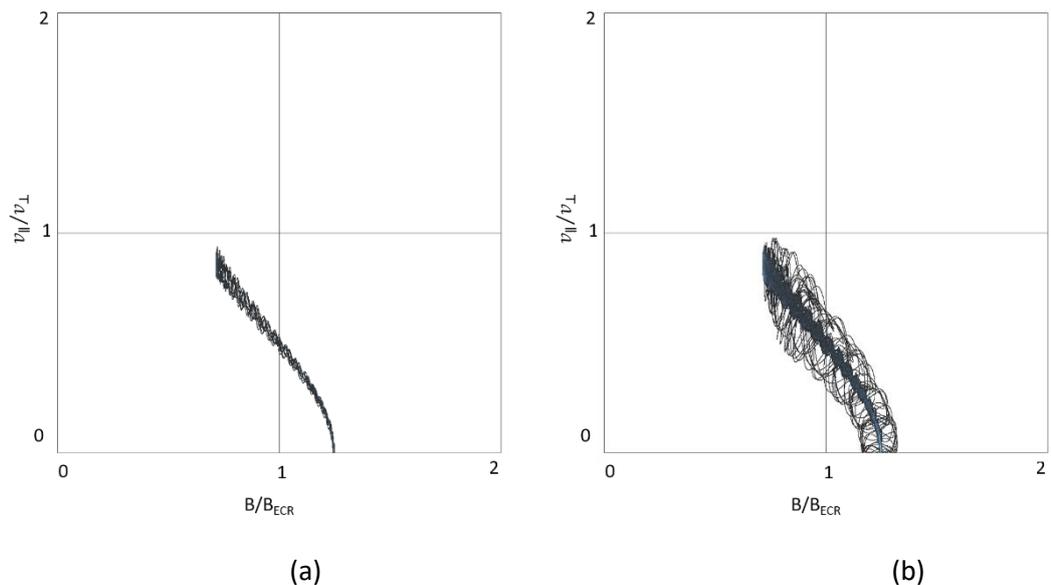

Fig.10. Trajectories of single electron in the space of "normalized local magnetic field B/B$_{ECR}$ - normalized longitudinal velocity $v_\parallel/v_\perp$" for displacements of the electron initial position by δy=0.235 cm and 0.245 cm for (a) and (b) respectively.

The critical displacement depends on the electron energy and magnetic field configuration. For the electron energy of 250 keV and the launching positions that are located at the corresponding ECR magnetic field of 0.777 T, the critical displacement decreases from 2.35 to 1.95 mm. For the minimum-B value of 0.92, the critical displacements are 3.25 mm and 2.65 mm for electron energies of 125 and 250 keV respectively; the minimum-B value substantially influences the critical displacements and subsequently the plasma core transversal size.

The radial magnetic confinement of electrons is relatively weaker than the axial one. In Fig.10, we visualize the electron movement by using local magnetic field normalized to the cold ECR field at the given point of the electron trajectory as the abscissa and, as the ordinate, ratio between longitudinal and perpendicular velocities of the particle in respect to the magnetic field line. The trajectories are shown for two launching positions of electron corresponding to the trajectories in Fig.9 (b) and (c). Points with the zero longitudinal velocities correspond to the positions of the electron, where it is longitudinally reflected back along the magnetic field line in the increasing magnetic field. The finite Larmor radius effect manifests itself as the oscillations of the velocity ratio at the relatively low magnetic fields close to the source center. For the axially confined electrons, spread in the turning (maximal) magnetic fields and in the corresponding turning positions is small. For the radially confined electrons, the spread is large, which results in shifts of the turning positions depending on the phase of the electron Larmor rotation at these points. Oscillations in separation of the turning points from the walls make the electron losses more probable and decrease the electron life time. Spread in the turning positions is proportional to the Larmor radius of electron, and therefore contribution to the electron loss rate due to this effect increases with the electron energy.

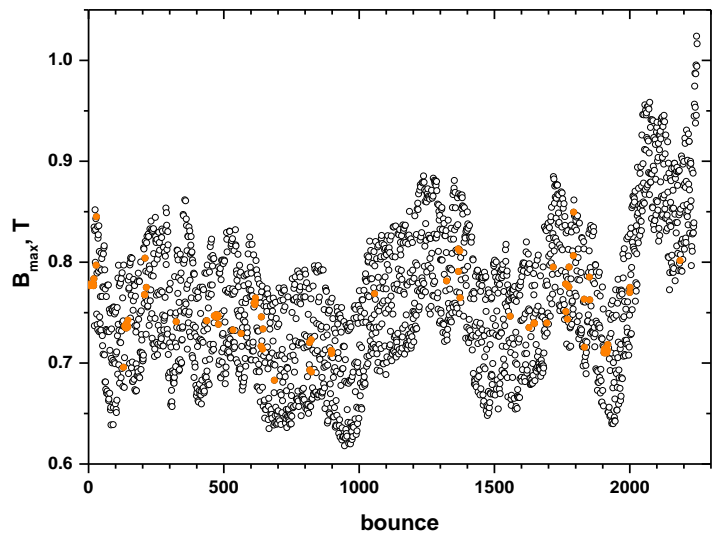

Fig.11. The maximal magnetic field at a single bounce as the function of the bounce number for the 250-keV electron with displacement of initial position of 0.197 cm for the magnetic configuration with $B_{min,n}$=0.715.

In addition to oscillations in the maximal magnetic fields due to finite Larmor radius, the nonadiabatic diffusion of magnetic moment facilitates the electron losses. In Fig.11, the turning (maximal) magnetic fields are shown as a function of the bounce number for 250-keV electron launched with the initial displacement of guiding center of 0.197 cm in the magnetic configuration with $B_{min,n}$=0.715. Electron orbit is unstable against transport out of the core – motion is switching to the radial mode in just 20 bounces. Moving along the magnetic field lines and experiencing the curvature and nonadiabatic particle drifts, the particle is radially lost after 2300 bounces, which corresponds to 1.4 μs of the particle movement time. Some of the turning positions are located close to the axis – we label these points with orange in Fig.11. The points are selected by requiring the relatively long time for a bounce, which corresponds to large distance between the turning points only available for movement close to the axis. Electron orbits are restoring to the axial mode from time to time, and then again return to the radial mode. Nonadiabatic diffusion affects the electron losses also in the axial direction as indicated by scatter of the turning magnetic fields close to the axis.

In Table 1, we list the data that are obtained by analyzing the electron trajectories for the electron energies of 1, 125 and 250 keV and for the investigated set of minimum-B values. For the low electron energies of 1 keV, transition from the axial to the radial motion is not so pronounced as for the large energies: the critical displacements for the 1-keV electrons correspond to change of the azimuthal drift direction from anti- to clockwise, with no substantial increase in the radial displacement of electrons from the source axis. In the Table, we also give the initial velocities of electrons normalized to the speed of light, the magnetic field gradients along the axis at the axis and at the positions of the ECR resonance for given electron energy. The longitudinal sizes along the axis of the ECR zone are shown, as well as time intervals between the zone crossing for those electrons that move along the axis. The data are used for estimations of the times for diffusion of electrons from the core and of the electron heating rates in the core.

Table 1. Parameters for estimation of the electron confinement and heating in the core plasma for different electron energies and minimum-B values.

| $B_{min,n}$ | 0.715 | | | 0.82 | | | 0.92 | | |
|---|---|---|---|---|---|---|---|---|---|
| E, keV | 1 | 125 | 250 | 1 | 125 | 250 | 1 | 125 | 250 |
| $B_{turn}$, T | 0.519 | 0.645 | 0.777 | 0.519 | 0.645 | 0.777 | 0.519 | 0.645 | 0.777 |
| $v_0/c$ | 0.0625 | 0.595 | 0.741 | 0.0625 | 0.595 | 0.741 | 0.0625 | 0.595 | 0.741 |
| $\alpha$, cm$^{-1}$ | 0.139 | 0.153 | 0.142 | 0.106 | 0.135 | 0.133 | 0.0616 | 0.118 | 0.125 |
| $\Delta z$, cm | 9.29 | 12.16 | 14.53 | 7.98 | 11.46 | 14.08 | 5.77 | 10.52 | 13.45 |
| $\tau_B$, ns | 14.05 | 1.58 | 1.39 | 15.1 | 1.65 | 1.43 | 16.8 | 1.70 | 1.47 |
| $R_c$, cm | 0.275 | 0.235 | 0.195 | 0.30 | 0.285 | 0.22 | 0.36 | 0.325 | 0.265 |
| $\tau_\perp$, arb. units | 1 | 2.9 | 3.1 | 0.89 | 3.78 | 3.72 | 0.69 | 4.23 | 5.11 |
| heating rate, arb. units | 1 | 0.35 | 0.28 | 1.34 | 0.39 | 0.3 | 2.48 | 0.41 | 0.32 |

The transversal diffusion of electrons in the core plasma is mostly defined by resonant interaction of electrons with the microwaves [10]. After crossing the resonance, electron gets a random kick $\Delta v_\perp$ in the transversal direction. Then, the Larmor radius is changed according to the kick magnitude and the local magnetic field, and the guiding center of electron orbits moves in random direction with the step of $\delta r \sim \gamma \Delta v_\perp / B$. As for the classical collisional diffusion, the electron transversal confinement time inside the plasma core is proportional $\tau_\perp \sim (R_c^2/(\delta r)^2)\tau_b$, where $\tau_b$ is the time interval between crossing the resonance and $R_c$ is the critical displacement of the electron orbit guiding center from the axis, at which the electron confinement switches to radial and electrons leave the core.

The time interval between the resonance crossings depends on spatial separation of the ECR points along the corresponding magnetic field lines and on the longitudinal velocities of electrons averaged over the bouncing cycle. By following the electron trajectories, we calculate these time intervals as $\tau_b$=14.05, 1.58 and 1.39 nsec for the electron energies of 1, 125 and 250 keV in the case of $B_{min,n}$=0.715, and 16.8, 1.7 and 1.47 nsec for $B_{min,n}$=0.92 (the relative increases are 1.2, 1.08, 1.06). With the increased minimum-B value, the resonant zone length along the source axis is decreasing, but this is overcompensated by decrease in the average longitudinal velocity. From adiabatic invariance of the magnetic moment $\mu = mv_\perp^2/2B$, it follows that the electron longitudinal velocity depends on the local magnetic field (B) as $v_\parallel = v_0\sqrt{1 - B/B_{turn}}$, and the larger is the minimum-B value, the slower is the electron movement between the turning points. The result is that the bouncing frequency decreases when $B_{min,n}$ increases.

The kick magnitude $\Delta v_\perp$ depends on the electron velocity as it is estimated in [20]: if electron passes through the resonance with the relatively constant velocity, then $\Delta v_\perp \sim \frac{e}{m\gamma}E_0\tau_{ECR} = \frac{e}{m\gamma}E_0 \times 1.13(\frac{2}{\alpha\omega_{RF}})^{1/2}v_\parallel^{-1/2}$, if electron starts to turn around at the resonance, $\Delta v_\perp \sim \frac{e}{m\gamma}E_0 \times \frac{0.71}{\omega_{RF}}(\frac{2\omega_{RF}}{\alpha v_\perp})^{2/3}$. Here, $\tau_{ECR}$ is the time for particle to pass through the resonance, $\omega_{RF}$ is the angular frequency of the microwaves, $\alpha = B_s^{-1}\frac{dB_s}{ds}$ is the normalized gradient of the magnetic field along the magnetic field line, and $E_0$ is local amplitude of the microwave electric field. For the electron trajectories depicted in Fig.9, the second estimation of the velocity kick is applicable. Then, we can calculate the transversal confinement times for those core electrons that turn around at the corresponding resonant magnetic fields, for different minimum-B values and for set of electron energies of 1, 125 and 250 keV. At this, we neglect possible variations of the microwave electric field amplitude for different magnetic fields and locations of the relativistic ECR zones. The times are normalized to the value for 1-keV electrons in the $B_{min,n}$=0.715 magnetic field configuration and are listed in Table 1.

Electron confinement time in the core is increasing with the electron energy mostly due to decrease in the velocity kick amplitudes. The longest confinement time is observed for the minimum-B value of 0.92, increasing by factor 1.65 in comparison to the value for $B_{min,n}$=0.715 because of the large critical displacement.

The electron heating rates are estimated for the parameters in Table 1 using the relation $dE/dt \sim \gamma^3/(\gamma^2 + \gamma)(\Delta v_\perp)^2/\tau_b$. Again, the rates are normalized to the value for 1 keV electrons with $B_{min,n}$=0.715. The heating rates are decreasing with the electron energy due to decrease in the velocity kick amplitudes. The largest heating rate for cold electrons is calculated for the large minimum-B value, increasing by factor of 2.5 compared to the case of the small $B_{min,n}$. The reason for such increase is that the magnetic field gradient close to the cold ECR field is lower by factor of almost two for $B_{min,n}$=0.92. Changes in the heating rates of energetic electrons for different minimum-B values are not significant: with the increased length of the relativistic ECR zone, the magnetic field gradients, as well as the bouncing times, become to be close each other.

Values of the heating rates and of the confinement times in Table 1 strongly depend on estimations of the velocity kicks and on assumption of fixed amplitude of the microwave electric field $E_0$. To check these estimations, we directly calculate the electron heating with microwaves in NAM-ECRIS using the spatial distributions of the microwave field amplitude and phase, and by tracing the electron movement in the magnetic trap. For heating of the core electrons with energies in the range (0.75-1.25) keV, we obtain the rates of 37±19 and 105±34 eV/μs for the $B_{min,n}$=0.715 and 0.92 respectively, with increase in the heating rate for the large minimum-B value close to estimations in Table 1. For the core electrons with energies in the range (100-150) keV, the heating rates differ not much from each other, specifically 90±30 and 60±20 eV/μs. These values are much higher than it is expected from the scaling given in Table 1, mostly because the scaling neglects the microwave absorption in the plasma; nonetheless, the global tendency of having the same energetic electron heating rates for different minimum-B values is confirmed by the direct calculations.

Concluding, we note that 1) the dense plasma core in ECRIS is formed due to the electrostatic trapping of the cold electrons along the source axis by the potentials at the biased electrode and at the extraction aperture, as well as by relatively good transversal magnetic confinement; 2) transversal diffusion of electrons from the core is dependent on the transition from predominantly axial to the radial movement of electrons governed by the hexapole component of the source magnetic field; 3) increase in the minimum-B magnetic field slows down the transversal diffusion and increases the energy confinement time for hot electrons in the core. The result is that electron population is substantially hotter for the large minimum-B values, in correspondence to the data in Fig.5, even if the heating rates for energetic electrons are comparable.

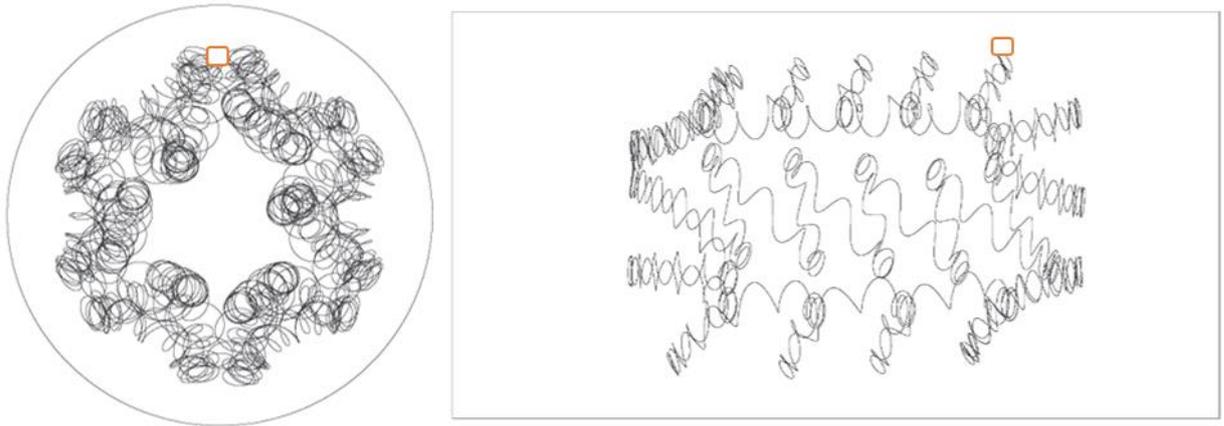

Fig.12 Trajectories of 250-keV electron in transversal (a) and longitudinal (b) planes for movement in the halo at $B_{min,n}$=0.715.

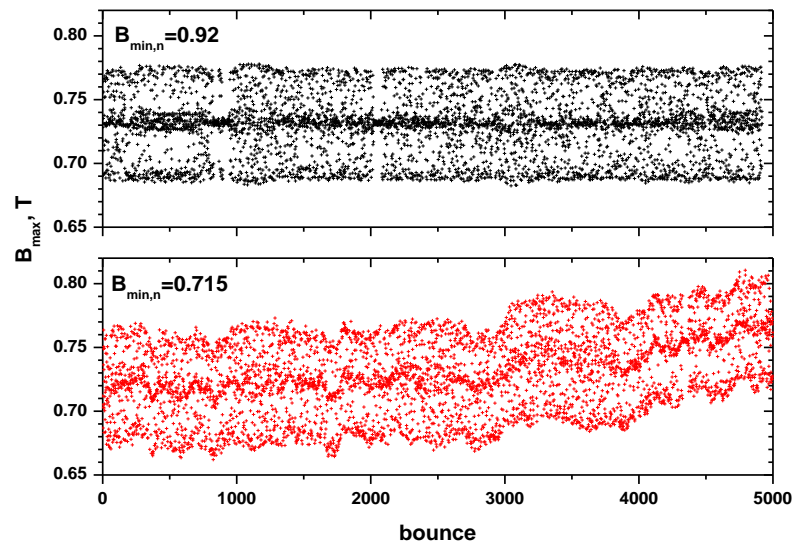

Fig.13. Maximal magnetic field for a single bounce as a function of bounce number for 150-keV electron at the magnetic field configurations with $B_{min,n}$=0.92 (black) and 0.715 (red).

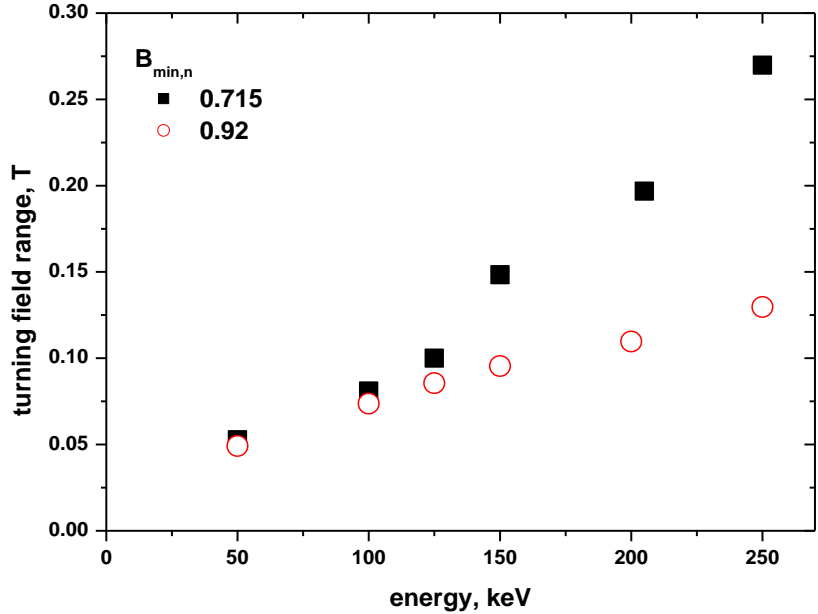

Fig.14. Ranges of turning magnetic fields in a dataset of 5000 bounces as a function of electron energy for $B_{min,n}$=0.715 (black squares) and 0.92 (open red circles).

As it is seen from Fig.11, movement of energetic electrons in the plasma halo is strongly affected by the nonadiabatic effects. To see details of electron dynamics in the halo and investigate the energy dependence of the nonadiabatic diffusion, we trace the single electron movement by launching it at the point, where the magnetic field is equal to 0.75 T at z-position of 17.7 cm, close to the leak position at radial wall (Fig.7). The particle is launched with zero longitudinal velocity and with transversal velocity that corresponds to energy varied in the range from 50 to 250 keV. In Fig.12, there are shown typical trajectories of such electron in transversal and longitudinal planes for electron energy of 250 keV and for $B_{min,n}$=0.715. The launching position is shown as the orange square at the plots.

The particle is bouncing along the magnetic field lines and is reflected by mirror effect always close to the radial walls. Its azimuthal drift is governed by the hexapole component of the magnetic field. Similar to what is shown in Fig.10, the turning magnetic fields oscillate due to finite Larmor radius. The nonadiabatic effects cause diffusion of the electron magnetic moment and of the turning magnetic fields; the diffusion rate depends on the magnetic field configuration and electron energy: generally, configurations with the relatively low minimum-B value are characterized by larger Larmor radii and larger gradients of the magnetic field (Fig.8), which makes the nonadiabatic processes stronger.

In Fig.13, the maximal magnetic fields are plotted as function of bounce number for single electron with energy of 150 keV and for the minimum-B values of 0.92 and 0.715. In total, 5000 bounces are traced, which corresponds to the physical time of electron movement of ~2 μsec for the given electron energy. Diffusion of the maximal magnetic fields is pronounced for the low minimum-B value, while motion of 150-keV electrons for the large minimum-B is stable. Whenever the particle's turning magnetic field exceeds the field on the wall, electron is lost.

It is important to note that diffusion of the turning magnetic fields results in the electron losses even in absence of particle collisions and interaction with microwaves. The faster is diffusion, the faster are the electron nonadiabatic losses. In Fig.14, we compare the diffusion rates for different electron energies and two minimum-B values of 0.715 and 0.92 by plotting the ranges of the turning magnetic fields – the difference between maximal and minimal values in dataset of 5000 points. For the stable motion at $B_{min,n}$=0.92, the range is slowly increasing with energy following the increased Larmor radius. For the low minimum-B value, substantial increase in the diffusion rate is seen for electrons with energies above 125 keV; for electron energy of 250 keV, the range is larger by factor of ~2.5 compared to the stable trajectories at $B_{min,n}$=0.92.The electron nonadiabatic losses are increasing with electron energies, which explains the formation of the bumps in the lost electron energy spectra shown in Fig.5. Larger minimum-B values suppress these losses and shift their energy threshold to higher values, which results in hotter electron population not only in the core, but also in the halo.

Influence of the electron nonadiabatic losses on the ECRIS performance did not attract much attention till now. However, the losses can be quite important and should be considered when analyzing the source responses to changes in magnetic field configurations. Dumping of the losses occurs with high magnetic fields and large dimensions of the source chamber. This partially explains why electrons are so hot in the sources with high microwave frequencies [4]. The plasma micro-instabilities at large minimum-B values [5] directly affect the extracted ion currents by modulating the particle life times and the plasma potential; increased rates for excitation of such instabilities can be due to increased anisotropic energy content when the nonadiabatic electron losses are dumped. The dense core plasma is confined in the regions with concave magnetic field lines, which is a potential source for excitation of macro-instabilities in overheated dense ECRIS plasma. Our model does not include impact of these instabilities; nonetheless, drop in the extracted ion currents in the large minimum-B configurations is reproduced.

## Acknowledgements

This work was supported by the Russian Foundation for Basic Research under grant No. 20-52-53026/20.

## References

[1] R. Geller, "Electron Cyclotron Resonance Ion Sources and ECR Plasma", (Institute of Physics, Bristol), 1996
[2] Hideyuki Arai, Masashi Imanaka, Sang-Moo Lee, Yoshihide Higurashi, Takahide Nakagawa, Masanori Kidera, Tadashi Kageyama, Masayuki Kase, Yasushige Yano, Toshimitsu Aihara, "Effect of minimum strength of mirror magnetic field (Bmin) on production of highly charged heavy ions from RIKEN liquid-He-free super conducting electron–cyclotron resonance ion source (RAMSES)", Nuclear Instruments and Methods in Physics Research Section A: Accelerators, Spectrometers, Detectors and Associated Equipment, **491**, 9 (2002); https://doi.org/10.1016/S0168-9002(02)01129-4
[3] J. B. Li, L. X. Li, B. S. Bhaskar, V. Toivanen, O. Tarvainen, D. Hitz, L. B. Li, W. Lu, H. Koivisto, T. Thuillier, J. W. Guo, X. Z. Zhang, H. Y. Zhao, L. T. Sun and H. W. Zhao, "Effects of magnetic configuration on hot electrons in a minimum-B ECR plasma", Plasma Phys. Control. Fusion, **62**, 095015 (2020); https://doi.org/10.1088/1361-6587/ab9d8f
[4] J. Benitez, C. Lyneis, L. Phair, D. Todd, D. Xie, "Dependence of the Bremsstrahlung Spectral Temperature in Minimum-B Electron Cyclotron Resonance Ion Sources", IEEE Transactions on Plasma Science, **45**,1746 (2017); https://doi.org/10.1109/TPS.2017.2706718
[5] O. Tarvainen, I. Izotov, D. Mansfeld, V. Skalyga, S. Golubev, T. Kalvas, H. Koivisto, J. Komppula, R. Kronholm, J. Laulainen and V. Toivanen, "Beam current oscillations driven by cyclotron instabilities in a minimum-B electron cyclotron resonance ion source plasma", Plasma Sources Sci. Technol., **23**, 025020 (2014). https://doi.org/10.1088/0963-0252/23/2/025020
[6] I. Izotov, O. Tarvainen, V. Skalyga, D. Mansfeld, H. Koivisto, R. Kronholm, V. Toivanen, and V. Mironov, "Measurements of the energy distribution of electrons lost from the minimum B-field—The effect of instabilities and two-frequency heating", Rev. Sci. Instrum., **91**, 013502 (2020); https://doi.org/10.1063/1.5128322
[7] B. Isherwood, G. Machicoane, "Measurement of the energy distribution of electrons escaping confinement from an electron cyclotron resonance ion source", Rev. Sci. Instrum., **91**, 025104 (2020); https://doi.org/10.1063/1.5129656
[8] V. Mironov, S. Bogomolov, A. Bondarchenko, A. Efremov, V. Loginov and D Pugachev, "Spatial distributions of plasma potential and density in electron cyclotron resonance ion source", Plasma Sources Sci. Technol., **29**, 065010 (2020); https://doi.org/10.1088/1361-6595/ab62dc
[9] I. Izotov, V. Skalyga, O. Tarvainen, E. Gospodchikov, A. Shalashov, H. Koivisto, R. Kronholm, V. Toivanen, V. Mironov, B. Bhaskar, "On the role of rf-scattering in the electron losses from minimum-B ECR plasmas", unpublished, arXiv:1912.04285v2 [physics.plasm-ph]
[10] V. Mironov, S. Bogomolov, A. Bondarchenko, A. Efremov, V. Loginov and D. Pugachev, "Three-dimensional modelling of processes in Electron Cyclotron Resonance Ion Source", JINST, **15**, P10030 (2020); https://doi.org/10.1088/1748-0221/15/10/P10030

[11] S. L. Bogomolov, A. E. Bondarchenko, A. A. Efremov, K. I. Kuzmenkov, A. N. Lebedev, V. E. Mironov, V. N. Loginov, N. Yu. Yazvitsky, N. N. Konev, "Production of High-Intensity Ion Beams from the DECRIS-PM-14 ECR Ion Source", Phys. Part. Nuclei Lett., **15**, 878 (2018); https://doi.org/10.1134/S1547477118070191

[12] P. Bellan, "Fundamentals of Plasma Physics", (Cambridge University Press, Cambridge), 2006; doi:10.1017/CBO9780511807183

[13] K. Halbach and R. F. Holsinger, "SUPERFISH - A Computer Program for Evaluation of RF Cavities with Cylindrical Symmetry", Particle Accelerators, **7**, 213 (1976); https://laacg.lanl.gov/laacg/services/sfregis.php

[14] COMSOL, Inc., COMSOL Multiphysics®, version 5.4, (2020) https://www.comsol.com/

[15] C. Lyneis, D. Leitner, D. Todd, S. Virostek, T. Loew, A. Heinen, and O. Tarvainen, "Measurements of bremsstrahlung production and x-ray cryostat heating in VENUS", Rev. Sci. Instrum., **77**, 03A342 (2006); https://doi.org/10.1063/1.2163870

[16] J. Noland, J. Y. Benitez, D. Leitner, C. Lyneis, and J. Verboncoeur, "Measurement of radial and axial high energy x-ray spectra in electron cyclotron resonance ion source plasmas", Rev.Sci. Instrum., **81**, 02A308 (2010); https://doi.org /10.1063/1.3258614

[17] T. Thuillier, J. Angot, J. Y. Benitez, A. Hodgkinson, C. M. Lyneis, D. S. Todd, and D. Z. Xie, "Investigation on the electron flux to the wall in the VENUS ion source", Rev. Sci. Instrum., **87**, 02A736 (2016); https://doi.org/10.1063/1.4935989

[18] B. V. Chirikov, "Resonance processes in magnetic traps", J. Nucl. Energy, Part C Plasma Phys., **1**, 253 (1960); https://doi.org/10.1088/0368-3281/1/4/311

[19] J.R. Roth, "Nonadiabatic Particle Losses in Axisymmetric and Multipolar Magnetic Fields", NASA TECHNICAL NOTE, NASA TN D-3164 (1965); https://ntrs.nasa.gov/api/citations/19660003733/downloads/19660003733.pdf

[20] M.A. Lieberman and A.J. Lichtenberg, "Theory of electron cyclotron resonance heating. II. Long time and stochastic effects", Plasma Phys., **15**, 125 (1973); http://dx.doi.org/10.1088/0032-1028/15/2/006